\documentclass[aps,prl,showkeys,twocolumn]{revtex4}

\usepackage{amsmath}
\usepackage{amssymb}
\usepackage{graphicx}
\usepackage{epsfig}         	

\newcommand{\beq}{\begin{equation}}
\newcommand{\eeq}{\end{equation}}
\newcommand{\bef}{\begin{figure}}
\newcommand{\eef}{\end{figure}}

\begin{document}

\title{An electrostatic model for biological cell division}
\author{Eshel Faraggi}
\email[Electronic address: 
]{faraggi@physics.utexas.edu}

\affiliation{Research and Information Systems, 155 Audubon Dr., Carmel, Indiana, 46032 USA}
\date{June 20, 2006}

\begin{abstract}
Probably the most fundamental processes for biological systems is their ability to create themselves through the use of cell division and cell differentiation. In this work a simple physical model is proposed for biological cell division. The model consists of a positive ionic gradient across the cell membrane, and concentration of charge at the nodes of the spindle and on the chromosomes. A simple calculation, based on Coulomb's Law, shows that under such circumstances a chromosome will tend to break up to its constituent  chromatids and that the chromatids will be separated by a distance that is an order of thirty percent of the distance between the spindle nodes. Further repulsion between the nodes will tend to stretch the cell and eventually break the cell membrane between the separated chromatids, leading to cell division. The importance of this work is in continuing the understanding of the electromagnetic basis of cell division and providing it with an analytical model. A central implication of this and other studies is to give theoretical support to the notion that cell division can be manipulated by electromagnetic means. Requirements on the ingredients of more sophisticated models for biological cell division will also discussed. {\em Copyright 2006--2010 by the author.}

\end{abstract}


\maketitle

\section{Introduction}
Understanding mitosis and  cytokinesis, the processes by which biological cells divide, has profound implications to practically all areas of medicine and biology. It also exposes the very important physical question: how does a complex, self organized, animate structure develop from mechanistic-inanimate constituents. An understanding of cell division and its manipulation could lead, e.g., to the ability to inhibit the growth of tumors, the ability to restart growth in defective or damaged cells, as well as many other exciting prospects. Unfortunately, though there has been significant progress in understanding cell division's mechanical parts,~\cite{mitbook,glot97,gagl02} a clear picture of the major underlying physical mechanisms involved does not exist. Such an understanding is the most efficient method to harness the potential in manipulating cell division.

In this report a simple mechanistic approach is presented for cell division. In this approach it is assumed that a charge gradient is created across the cell membrane through protein interactions and that positive ions such as ${Ca}^{+2}$ and ${Na}^+$ enter through these ion channels into the cell's spindle structure,  a network of protein based tubes underlying the innards of the cell during cell division. Though this is a feasible construct~\cite{mitosis}, it is not claimed here to be the actual mechanism of cell division but rather as a test for the required electromagnetic interactions. Out of the four fundamental forces known in nature it is only the electromagnetic force that can play a major role in small scale biology in general and cell division in particular. Charge gradients play a significant role in other biological systems such as nerves.

\section*{Theory}
\begin{figure}[!ht]
\begin{center}
\includegraphics[width=3in]{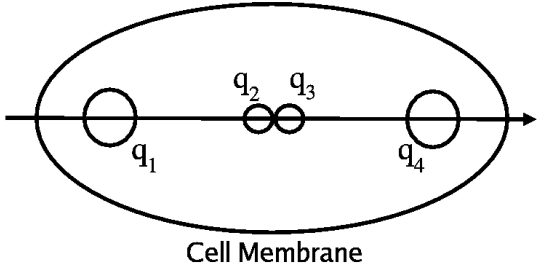}
\end{center}
\caption{
{\bf Components of the minimal cell.}  A cell membrane, two spindle centers, $q_1$ and $q_4$, and two chromatids, $q2$ and $q3$, which carry the genetic information. Figure not to scale.
}
\label{simpcel}
\end{figure}

Let us discuss the potential energy of the system. To allow the model to be transparent, only the bare minimum cell components are included: a cell membrane, two spindle centers with charge $q_1=q_4$, and two chromatids with charge $q_2=q_3$. To a given charge $q_i$, with $i=1,2,3,4$, a one dimensional position $x_i(t)$ and velocity $v_i(t)$ are associated. The coordinate system is chosen such that, $x_1=-x_4$ and $x_2=-x_3$ initially. A diagram of this configuration is given in Fig.~\ref{simpcel}. The initial velocity is taken to be zero for all four charges. These assumptions will insure that $x_1=-x_4$ and $x_2=-x_3$ for all times. The rational for choosing to work on a one-dimensional problem goes beyond the attempt to present the most transparent model. The structure of the dividing cell would define a preferred spatial direction along the line joining the two spindle centers, hence a one-dimensional model is appropriate for the position of the four charges.

The classical electrostatic contribution to the energy can be expressed as
\beq
\label{emener}
E_{ES} = \frac{1}{D} (\frac{q_1^2}{l}+\frac{q_2^2}{d} + 8 q_1 q_2  \frac{l}{l^2-d^2}),
\eeq
with $l=x_4-x_1$, $d=x_3-x_2$, and $D$ is the inverse dielectric constant. Here it will be assumed that the cell can be modeled as water and $D=80$ is taken. The symmetry of the system about the origin was used to arrive at Eq.~(\ref{emener}). The fact that the electrostatic potential energy is positive relative to the charges at infinity comes about from the repulsive interaction between the positive charges, since the system would prefer to be in a lower energy or a more separated state.

A well known result for the electrostatic field inside biological media and other neutral solution is the Debye length. This calculation shows that for a neutral solution with free positive and negative ions distributed uniformly, the electrostatic field due to an ion drops off exponentially once the field it generates is in the vicinity of opposite charges. The reason the notion of a Debye length does not invalidate the model proposed here is that the charge distribution in the dividing cell is not neutral, nor does the mitotic spindle allow the homogeneous distribution of ions throughout the cell. It has been known for several decades that charge pumping into the cell and mitotic spindle accompanies cell division. (see for example Ref.~\cite{mitosis} p. 263). Hence, The charge density in the dividing cell is not properly accounted for under the assumption of a neutral solution. One should also remember that the existence of positive charges in the cell's media does not reduce the electrostatic repulsion force between charges $q_1$ and $q_4$ in Fig.~\ref{simpcel}, it increases it.

At this point it is instructive to find the minima of the electrostatic energy, as this will illuminate on how the charged chromatids separate the genetic code of the cell. If we assume for simplicity that $l_0=10 \, \mu m$ and constant, the minimum of Eq.~(\ref{emener}) is given by the implicit condition for $d_0$, $1/d_0^2 = 16 d_0 l_0 / (l_0^2-d_0^2)^2$, for $q_1=q_2$. This condition can be solved to give $d_0=3.6148 \, \mu m$, i.e., if we start with a configuration of chromatids that is close together they will tend to separate to a distance that is of the order of magnitude of the initial cell size. Of course based on Eq.~(\ref{emener}) $l$ itself would tend to increase as well, which will finalize the separation of the cell. However, as we shall soon see changes in $l$ occur on a much larger time scale than changes in $d$.

The next ingredient in this presentation is the deformation of the cell membrane. This is a crucial step both for the initial division of the cell, and equally important, in the capability of the daughter cells to separate from each other. Here the approach is to assume that the cell membrane is a linear catastrophic spring, i.e., that it resists stretching by a force that is proportional to its perturbation in length and that given some catastrophic condition the spring will break, and the force between its two ends will be zero, representing the separation of the two cell membranes. This force is applied to the spindle centers, $q_1, \, q_4$  but not to the chromatids, $q_2, \, q_3$, which are assumed to be small in comparison to the cell size, free of the tensions of the cell membrane, and approximately not affected by the  nuclear membrane. The potential energy associated with a linear spring can be written as
\beq
\label{stenrgy}
E_{ST} = \frac{1}{2} k (x_1(t)-x_1(0))^2 + \frac{1}{2} k (x_4(t)-x_4(0))^2 ,
\eeq
with $k$ the spring constant.

The catastrophic condition is taken to be a stretching of the cell membrane to twice its length. When this happens it is assumed that $E_{ST}=0$. Under the assumptions presented so far, maximum stress will be experienced by the cell membrane at the position $x=0$. Hence it will be assumed that the cell membrane will break at $x=0$. The actual separation of the two daughter cells involves joining of opposing cell membranes around $x=0$.

The movement of the chromatids inside the cell is important to the process of separation. The details of the separation of a chromosome into its two chromatids probably involves the breaking of molecular bonds, for example by charge substitution. However for the rest it will be assumed that the details of separation of the chromatids can be modeled as a separate process and will be ignored.

The details of the various forces the chromatids experience in the process of division are again relatively complex. The drag force acting on the chromatids, i.e., on $q_2$ and $q_3$, is
\beq
F_{d,i} = -\gamma v_i(t)
\label{dragforce}
\eeq
with $i=2,3$, $\gamma$ is a dissipative parameter. As shall be shown shortly, this parameter is important to the behavior of the chromatids.

\section{Results}
We can get a relatively quick estimate for the number of charges involved in the process by considering a simpler two charge problem. Let us define the average force, $\bar{F}$, from the initial combined stage, $x_i$, and the final separated stage, $x_f$, 
\beq
\label{fbar}
\int_{x_i}^{x_f} \bar{F} dx = \int _{x_i}^{x_f} F dx = -(E_{ES}(x_f) - E_{ES}(x_i)),
\eeq
with $F=-\partial E_{ES} / \partial x$, including for the moment only the electrostatic interaction, and $x$ is chosen as the position of one of the charges. Now, we can associate with this average force an average acceleration, $\bar{a}$, obeying 
\beq
\label{dx.eq}
x_f - x_i = \frac{1}{2} \bar{a} (t_f - t_i)
\eeq
with $t_i$ and $t_f$ the initial and final times.

The change in the electrostatic potential energy for two identical charges with charge $q$ is given by $q^2 (1/x_f - 1/x_i)$. Combining with Eqs.~(\ref{fbar}) and~(\ref{dx.eq}) it is found that
\beq
\label{q.eq}
q^2 = \frac{2 (x_f-x_i) m x_i x_f}{(t_f - t_i)^2} D.
\eeq
with $m$ the accelerated mass. $m$ can be estimated as the mass of a sphere of water $10 \mu m$ in radius, $m = 10^{-8}  \, g$. If we use $x_i = 5 \, \mu m$, $x_f = 10 \, \mu m$, and $t_f - t_i = 10 \, s$ as a typical time scale for the separation phase, we find from Eq.~(\ref{q.eq}) a charge that is equivalent to about 10 $Ca^{+2}$ atoms. This order of magnitude estimate should be viewed as a feasibility test for the proposed theory. 

With the aide of a numerical simulation we can much better calculate the dynamics of the system. For simplicity it will be assumed that due to the structure of the spindle the two halves of the splitting cell can be represented by the mass of the dominant charges $m_1$ and $m_4$. The system is started with four charges at rest, situated at $-x_1(t=0) = x_4(t=0)=5 \, \mu$m,  $-x_2(t=0) = x_3(t=0)=10$~nm, with charge $q_1=q_4=10 e$, $q_2=q_3=2 e$, where $e$ is the charge of the proton. The main accelerated mass is composed of the charged ions but mostly of other cell materials contained in half the cell, $m_1=m_4=10^{-8}$~g. The mass of the chromatids is estimated as the mass of water occupying a tube of length ten $\mu m$ and radius ten $nm$, $m_2=m_3=10^{-15} \, g$. The parameter $k$ in Eq.~(\ref{stenrgy}) is estimated by the arbitrary requirement that at the catastrophic condition the force due to the surface tension would equal the electrostatic force between $q_1$ and $q_4$. Hence, $k = 10^{-10}$~g/s$^2$. 

Using these values and $\gamma=1^{-10}$~g/s for the parameters determining the system, the dynamic evolution is simulated. The system is started from the initial condition and an average acceleration is calculated, then it is evolved by a $\Delta t$, new positions and velocities are calculated using the calculated accelerations, etc. It was found that a time step of 1~ns produced convergence to within graphical resolution, i.e., reducing $\Delta t$ further did not change the graphical appearance of the solution.

\begin{figure}[!ht]
\begin{center}
\includegraphics[width=3in]{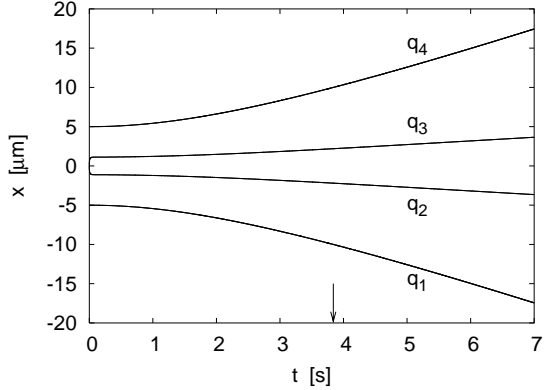}
\end{center}
\caption{
{\bf Position as a function of time for the system described in the text.}  
$-x_1(t=0) = x_4(t=0)=5 \, \mu$m,  $-x_2(t=0) = x_3(t=0)=10$~nm,  $q_1=q_4=10 e$, $q_2=q_3=2 e$, with $e$ the charge of the proton. $m_1=m_4=10^{-8}$~g, $m_2=m_3=10^{-15}$~g. $k = 10^{-8}$~g/s$^2$, $\gamma=1^{-10}$~g/s.
Both the cases of  $\Delta t = 1$~ns and $\Delta t =10$~ns are graphed and produce identical results to within graphical resolution.
}
\label{mitres}
\end{figure}

\begin{figure}[!ht]
\begin{center}
\includegraphics[width=3in]{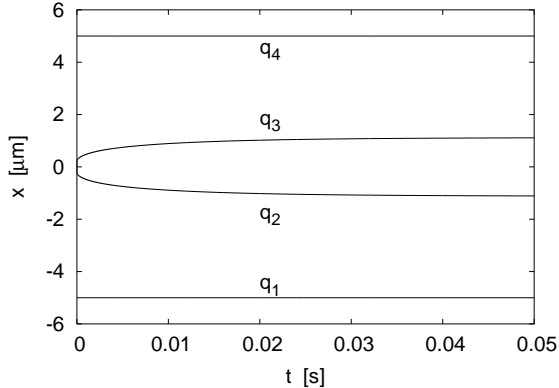}
\end{center}
\caption{
{\bf Blowup of Fig.~\ref{mitres}.} Used to show the earlier response of the chromatids. The drag parameter $\gamma$ will determine the type of relaxation the chromatids will experience. Both the cases of  $\Delta t = 1$~ns and $\Delta t = 10$~ns are graphed and produce identical results to within graphical resolution. 
}
\label{mitrezom}
\end{figure}

Results from such a calculation are presented in Fig.~\ref{mitres} and Fig.~\ref{mitrezom}. Since the aim of this calculation is to present a single mitotic event it was assumed that after the catastrophic event, when the cell has elongated to twice its original length, both $E_{ST}=0$ and $E_{ES}=0$. The time when this condition is imposed is indicated by an arrow in Fig.~\ref{mitres}. The differentiation of the original cell is clearly evident in Fig.~\ref{mitres}. Note that the original distance between the two chromatids is not visible on the scale of the figure, while the eventual separation of the chromatids is of the order of magnitude of the original cell size.

The initial details of the separation are presented in Fig.~\ref{mitrezom}, which is a blowup from Fig.~\ref{mitres}. It is evident from the figure that due to the relative closeness of the chromatids and their relatively low mass, their response times are much shorter than the those of the relatively massive poles. Hence, the dynamics in the initial stages will be dominated by that of the chromatids. This relates back to the previous conjecture that the chromatids act as initiators of cell division. Another important aspect to be gathered from Fig.~\ref{mitrezom} is that the parameter $\gamma$ plays a role in the movement of the chromatids and hence the dynamics of cell division. For low $\gamma$ the chromatids will undergo more oscillations than for high $\gamma$ where the chromatids slowly move to their separated configuration. The case presented in Fig.~\ref{mitrezom} is mostly steady relaxation though there is some overshoot in the separation between the chromatids.

\begin{figure}[!ht]
\begin{center}
\includegraphics[width=3in]{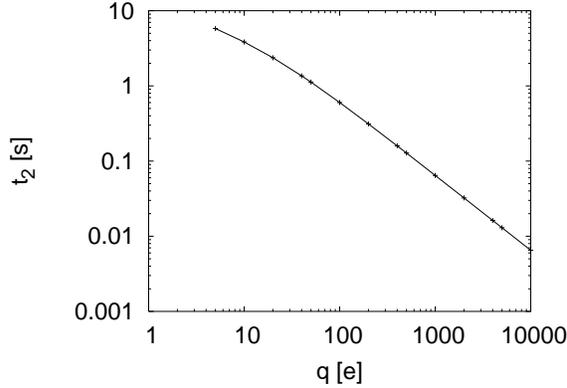}
\end{center}
\caption{
{\bf Time of separation.}  $t_2$, as a function of the charge of $q_1$ and $q_4$ which is denoted by $q$. For high $q$ values the charges $q_2$ and $q_3$ become negligible and $t_2 \propto 1/q$ as expected for two electrostatic charges. For low $q$ values the effects of the charge of $q_2$ and $q_3$ become more important and $t_2$ deviates from the $1/q$ dependence.
}
\label{t2vs.q}
\end{figure}

Finally, it is interesting to investigate the relationship between the time of separation and the number of charges  involved. The condition for separation that will be used here is that the distance between $q_1$ and $q_4$ is double its original distance. The time that it takes the cell to reach this point is denoted as $t_2$. In Fig.~\ref{t2vs.q}, $t_2$ is given as a function of the charge of $q_1$ and $q_4$ which is denoted by $q$. It should be observed that the for high $q$ the electrostatic effect from the charges $q_2$ and $q_3$ become negligible and $t_2 \propto 1/q$ as expected for two charges. For low $q$ the effects of the charges $q_2$ and $q_3$ become more important and $t_2$ deviates from $1/q$ dependency.

\section{Conclusions}
In summary, a transparent model was presented exhibiting cell division involving charge separation. It was shown that through the resulting electrostatic interaction the cell will separate into two symmetric parts, each  containing one pole and one chromatid.

Probably most important is that cell division is inherently an electromagnetic  phenomena. Fundamentally there is no other choice. Hence, it should be possible to manipulate cell division by electromagnetic fields. For example, in the treatment of cancer it may be possible to inhibit tumor growth by creating electromagnetic fields that disfavor cell division. The converse is also true, it may be possible to stimulate tissue growth by electromagnetic fields that favor cell division.

The presentation given here took the most basic approach, so as to be most transparent. Topics such as the regulation of charge content throughout the cell, which were mostly ignored here, are critical to the further understanding of and forthcoming observations about cell division.

After the formulation of theses ideas a related paper was found.~\cite{kirs04} In it the effects of high-frequency alternating electric fields on cancer tissue growth are studied. For frequencies in the range $100-300$~kHz it was found that  alternating electric fields have an inhibitory effect on the ability of the cancer cells to divide. Other evidence suggests that low frequency alternating electric fields with frequencies less than approximately $1$~kHz can promote tissue growth in some cases.~\cite{bass85} While not directly supporting the work presented here these studies strongly suggest the manipulation of cell division is possible with electromagnetic fields and this in turn confirms that cell division is inherently an electromagnetic phenomena.

\begin{acknowledgments}
The author would like to acknowledge the continuous support of Natali Teszler.

\end{acknowledgments}

\bibliography{mitosis}

\begin{thebibliography}{6}
\expandafter\ifx\csname natexlab\endcsname\relax\def\natexlab#1{#1}\fi
\expandafter\ifx\csname bibnamefont\endcsname\relax
  \def\bibnamefont#1{#1}\fi
\expandafter\ifx\csname bibfnamefont\endcsname\relax
  \def\bibfnamefont#1{#1}\fi
\expandafter\ifx\csname citenamefont\endcsname\relax
  \def\citenamefont#1{#1}\fi
\expandafter\ifx\csname url\endcsname\relax
  \def\url#1{\texttt{#1}}\fi
\expandafter\ifx\csname urlprefix\endcsname\relax\def\urlprefix{URL }\fi
\providecommand{\bibinfo}[2]{#2}
\providecommand{\eprint}[2][]{\url{#2}}

\bibitem[{\citenamefont{{J. S. Hyams and B. R. Brinkley}}(1989)}]{mitbook}
\bibinfo{editor}{\bibnamefont{{J. S. Hyams and B. R. Brinkley}}}, ed.,
  \emph{\bibinfo{title}{{MITOSIS: Molecules and Mechanisms}}}
  (\bibinfo{publisher}{Academic Press Inc.}, \bibinfo{year}{1989}).

\bibitem[{\citenamefont{Glotzer}(1997)}]{glot97}
\bibinfo{author}{\bibfnamefont{M.}~\bibnamefont{Glotzer}},
  \bibinfo{journal}{Curr. Opin. Cell. Biol.} \textbf{\bibinfo{volume}{9}},
  \bibinfo{pages}{815} (\bibinfo{year}{1997}).

\bibitem[{\citenamefont{Gagliardi}(2002)}]{gagl02}
\bibinfo{author}{\bibfnamefont{L.~J.} \bibnamefont{Gagliardi}},
  \bibinfo{journal}{Phys. Rev. E} \textbf{\bibinfo{volume}{66}},
  \bibinfo{pages}{011901} (\bibinfo{year}{2002}).

\bibitem[{\citenamefont{Hepler}(1989)}]{mitosis}
\bibinfo{author}{\bibfnamefont{P.~K.} \bibnamefont{Hepler}}, in
  \emph{\bibinfo{booktitle}{{MITOSIS: Molecules and Mechanisms}}}, edited by
  \bibinfo{editor}{\bibnamefont{{J. S. Hyams and B. R. Brinkley}}}
  (\bibinfo{publisher}{Academic Press Inc.}, \bibinfo{year}{1989}),
  chap.~\bibinfo{chapter}{7}, pp. \bibinfo{pages}{241--271}.

\bibitem[{\citenamefont{et~al.}(2004)}]{kirs04}
\bibinfo{author}{\bibfnamefont{E.~D.~K.} \bibnamefont{et~al.}},
  \bibinfo{journal}{Cancer Research} \textbf{\bibinfo{volume}{64}},
  \bibinfo{pages}{3288} (\bibinfo{year}{2004}).

\bibitem[{\citenamefont{Basset}(1985)}]{bass85}
\bibinfo{author}{\bibfnamefont{C.~A.} \bibnamefont{Basset}},
  \bibinfo{journal}{Clin. Plast. Surg.} \textbf{\bibinfo{volume}{12}},
  \bibinfo{pages}{259} (\bibinfo{year}{1985}).

\end{thebibliography}

\end{document}